\documentclass[3p,times,twocolumn]{elsarticle}

\usepackage{epsfig}
\usepackage{graphicx,subfigure}
\usepackage{lineno}
\usepackage[usenames]{color}
\biboptions{sort&compress}

\usepackage{ecrc}

\volume{00}

\firstpage{1}

\runauth{D.V. Poda et al.}

\usepackage{amssymb}
\usepackage[figuresright]{rotating}

\begin{document}

\begin{frontmatter}



\dochead{}

%---------------------------------------------------------------------------------

\title{Scintillating bolometers based on ZnMoO$_4$ and Zn$^{100}$MoO$_4$ crystals to search for 
0$\nu$2$\beta$ decay of $^{100}$Mo (LUMINEU project): first tests at the Modane Underground Laboratory}


\author[CSNSM,KINR,LUM]{D.V. Poda}
\ead{Denys.Poda@csnsm.in2p3.fr}
\fntext[LUM]{The LUMINEU Collaboration}
\fntext[EDW]{The EDELWEISS Collaboration}
\author[IRFU,EDW]{E.~Armengaud}
\author[IPNL,EDW]{Q.~Arnaud}
\author[IPNL,EDW]{C.~Augier}
\author[IPNL,EDW]{A.~Beno\^{\i}t}
\author[NEEL,EDW]{A.~Beno\^{\i}t}
\author[CSNSM,LUM,EDW]{L.~Berg\'e}
\author[KINR,LUM]{R.S.~Boiko}
\author[KIT-IPE,EDW]{T.~Bergmann}
\author[KIT-IEKP,KIT-IKP,EDW]{J.~Bl$\mbox{\"u}$mer}
\author[CSNSM,EDW]{A.~Broniatowski}
\author[JINR,EDW]{V.~Brudanin}
\author[NEEL,EDW]{P.~Camus}
\author[IPNL,EDW]{A.~Cazes}
\author[IPNL,EDW]{B.~Censier}
\author[CSNSM,LUM,EDW]{M.~Chapellier}
\author[IPNL,EDW]{F.~Charlieux}
\author[CSNSM,KINR,LUM]{D.M.~Chernyak}
\author[IAS,LUM]{N.~Coron}
\author[Oxford,EDW]{P.~Coulter}
\author[KIT-IEKP,EDW]{G.A.~Cox}
\author[KINR,LUM]{F.A.~Danevich}
\author[IRFU,EDW]{T.~de~Boissi\`{e}re}
\author[ICMCB,LUM]{R.~Decourt}
\author[IPNL,EDW]{M.~De~Jesus}
\author[Orphee,LUM]{L.~Devoyon}
\author[CSNSM,LUM,EDW]{A.-A.~Drillien}
\author[CSNSM,LUM,EDW]{L.~Dumoulin}
\author[KIT-IKP,EDW]{K.~Eitel}
\author[IAP,LUM]{C.~Enss}
\author[JINR,EDW]{D.~Filosofov}
\author[IAP,LUM]{A.~Fleischmann}
\author[IRFU,EDW]{N.~Fourches}
\author[IPNL,EDW]{J.~Gascon}
\author[IAP,LUM]{L.~Gastaldo}
\author[IRFU,EDW]{G.~Gerbier}
\author[CSNSM,Como,INFN-Bicocca,LUM]{A.~Giuliani}
\author[IRFU,LUM,EDW]{M.~Gros}
\author[KIT-IKP,EDW]{L.~Hehn}
\author[Oxford,EDW]{S.~Henry}
\author[IRFU,LUM,EDW]{S.~Herv\mbox{\'e}}
\author[KIT-IEKP,EDW]{G.~Heuermann}
\author[CSNSM,LUM,EDW]{V.~Humbert}
\author[NIIC,LUM]{I.M.~Ivanov}
\author[IPNL,EDW]{A.~Juillard}
\author[IPNL,KIT-IEKP,EDW]{C.~K\'ef\'elian}
\author[KIT-IPE,EDW]{M.~Kleifges}
\author[KIT-IEKP,EDW]{H.~Kluck}
\author[KINR,LUM]{V.V.~Kobychev}
\author[Orphee,LUM]{F.~Koskas}
\author[KIT-IKP,EDW]{V.~Kozlov}
\author[Oxford,EDW]{H.~Kraus}
\author[Sheffield,EDW]{V.A.~Kudryavtsev}
\author[CSNSM,EDW]{H.~Le~Sueur}
\author[LNAB,LUM]{M.~Loidl}
\author[IRFU,LUM]{P.~Magnier}
\author[NIIC,LUM]{E.P.~Makarov}
\author[CSNSM,DSATUI,LUM]{M.~Mancuso}
\author[CSNSM,LUM]{P.~de~Marcillac}
\author[CSNSM,LUM,EDW]{S.~Marnieros}
\author[CSNSM,LUM]{C.~Marrache-Kikuchi}
\author[KIT-IPE,EDW]{A.~Menshikov}
\author[NIIC,LUM]{S.G.~Nasonov}
\author[IRFU,LUM,EDW]{X-F.~Navick}
\author[IRFU,LUM,EDW]{C.~Nones}
\author[CSNSM,LUM,EDW]{E.~Olivieri}
\author[IRAMIS,EDW]{P.~Pari}
\author[IRFU,LUM,EDW]{B.~Paul}
\author[IRFU,LUM]{Y.~Penichot}
\author[INFN-Bicocca,DFUMB,LUM]{G.~Pessina}
\author[CSNSM,EDW]{M.C.~Piro}
\author[CSNSM,LUM]{O.~Plantevin}
\author[IAS,LUM]{T.~Redon}
\author[Sheffield,EDW]{M.~Robinson}
\author[LNAB,LUM]{M.~Rodrigues}
\author[JINR,EDW]{S.~Rozov}
\author[IPNL,EDW]{V.~Sanglard}
\author[KIT-IEKP,EDW]{B.~Schmidt}
\author[NIIC,LUM]{V.N.~Shlegel}
\author[KIT-IKP,EDW]{B.~Siebenborn}
\author[Orphee,LUM]{O.~Strazzer}
\author[KIT-IPE,EDW]{D.~Tcherniakhovski}
\author[CSNSM,LUM]{M.~Tenconi}
\author[IAS,LUM]{L.~Torres}
\author[KINR,LUM]{V.I.~Tretyak}
\author[IPNL,EDW]{L.~Vagneron}
\author[NIIC,LUM]{Ya.V.~Vasiliev}
\author[ICMCB,LUM]{M.~Velazquez}
\author[ICMCB,LUM]{O.~Viraphong}
\author[KIT-IKP,EDW]{R.J.~Walker}
\author[KIT-IPE,EDW]{M.~Weber}
\author[JINR,EDW]{E.~Yakushev}
\author[Oxford,EDW]{X.~Zhang}
\author[CML,LUM]{V.N.~Zhdankov}
\author[]{(for the LUMINEU and the EDELWEISS Collaborations)}

\address[CSNSM]{CSNSM, Centre de Sciences Nucl\'{e}aires et de Sciences de la Mati\`{e}re, CNRS/IN2P3, Universit\'e Paris-Sud, 91405 Orsay, France}
\address[KINR]{Institute for Nuclear Research, MSP 03680 Kyiv, Ukraine}
\address[IRFU]{CEA, Centre d'Etudes Saclay, IRFU, 91191 Gif-Sur-Yvette Cedex, France}
\address[IPNL]{IPNL, Universit\'{e} de Lyon, Universit\'{e} Lyon 1, CNRS/IN2P3, 69622 Villeurbanne Cedex, France}
\address[NEEL]{Institut N\'{e}el, CNRS/UJF, 38042 Grenoble Cedex 9, France}
\address[KIT-IPE]{Karlsruhe Institute of Technology, Institut f\"{u}r Prozessdatenverarbeitung und Elektronik, 76021 Karlsruhe, Germany}
\address[KIT-IEKP]{Karlsruhe Institute of Technology, Institut f\"{u}r Experimentelle Kernphysik, 76128 Karlsruhe, Germany}
\address[KIT-IKP]{Karlsruhe Institute of Technology, Institut f\"{u}r Kernphysik, 76021 Karlsruhe, Germany}
\address[JINR]{Laboratory of Nuclear Problems, JINR, 141980 Dubna, Moscow region, Russia}
\address[IAS]{IAS, CNRS, Universit\'{e} Paris-Sud, 91405 Orsay, France}
\address[Oxford]{University of Oxford, Department of Physics, Oxford OX1 3RH, UK}
\address[ICMCB]{ICMCB, CNRS, Universit\'{e} de Bordeaux, 33608 Pessac Cedex, France}
\address[Orphee]{CEA, Centre d'Etudes Saclay, Orph\'{e}e, 91191 Gif-Sur-Yvette Cedex, France} 
\address[IAP]{Institut f\"{u}r Angewandte Physik, Universit\"{a}t Heidelberg, D-69120 Heidelberg, Germany}
\address[Como]{Dipartimento di Scienza e Alta Tecnologia dell’Universit`a dell’Insubria, 22100 Como, Italy}
\address[INFN-Bicocca]{INFN, Sezione di Milano-Bicocca, 20126 Milano, Italy}
\address[NIIC]{Nikolaev Institute of Inorganic Chemistry, 630090 Novosibirsk, Russia}
\address[Sheffield]{Department of Physics and Astronomy, University of Sheffield, Hounsfield Road, Sheffield S3 7RH, UK}	
\address[LNAB]{CEA, Centre d'Etudes Saclay, LNAB, 91191 Gif-Sur-Yvette Cedex, France} 
\address[DSATUI]{Dipartimento di Scienza e Alta Tecnologia dell'Universit\'{a} dell'Insubria, I-22100 Como, Italy} 
\address[IRAMIS]{CEA, Centre d'Etudes Saclay, IRAMIS, 91191 Gif-Sur-Yvette Cedex, France} 
\address[DFUMB]{Dipartimento di Fisica dell'Universit\'{a} di Milano-Bicocca, 20126 Milano, Italy} 
\address[CML]{CML Ltd., 630090 Novosibirsk, Russia}

\begin{abstract}
The technology of scintillating bolometers based on zinc molybdate (ZnMoO$_4$) 
crystals is under development within the LUMINEU project to search for 
0$\nu$2$\beta$ decay of $^{100}$Mo with the goal to set the basis for large scale 
experiments capable to explore the inverted hierarchy region of the 
neutrino mass pattern. Advanced ZnMoO$_4$ crystal scintillators with mass 
of $\sim$~0.3 kg were developed and Zn$^{100}$MoO$_4$ crystal from enriched 
$^{100}$Mo was produced for the first time by using the low-thermal-gradient 
Czochralski technique. One ZnMoO$_4$ scintillator and two samples 
(59 g and 63 g) cut from the enriched boule were tested aboveground at 
milli-Kelvin temperature as scintillating bolometers showing a high 
detection performance. The first results of the low background measurements 
with three ZnMoO$_4$ and two enriched detectors installed in the EDELWEISS 
set-up at the Modane Underground Laboratory (France) are presented. 
\end{abstract}


\end{frontmatter}

%\linenumbers

\section{Introduction}

Scintillating bolometers --- cryogenic detectors with a heat-light double read-out --- can play 
a crucial role in next-generation experiments to study neutrino properties and weak interaction 
via investigating neutrinoless double beta (0$\nu$2$\beta$) decay, as discussed in Refs. 
\cite{Bee12,Art14}. This technique is extensively developing now within the 
LUCIFER \cite{Fer11,Bee13}, the AMoRE \cite{Bha12,Kim14}, and the LUMINEU \cite{Ten13} 
0$\nu$2$\beta$ projects. This paper describes the recent achievements in the framework of the 
LUMINEU programme (Luminescent Underground Molybdenum Investigation for NEUtrino mass and nature). 

LUMINEU is devoted to the development of a technology based on zinc molybdate (ZnMoO$_4$) 
scintillating bolometer as a basis for the realization of a high-sensitivity 0$\nu$2$\beta$ 
experiment. The good prospects of this material for the bolometric technique are clearly 
shown in recent investigations \cite{Gir10,Bee12,Bee12a,Bee12b,Bee12c,Che13,Ber14}. 
An important point in the realization of LUMINEU is concerned with the technology of growing 
high-quality radiopure large mass (0.3--0.5 kg) ZnMoO$_4$ single crystals 
with the aim to produce scintillators enriched in $^{100}$Mo (Zn$^{100}$MoO$_4$). 
Here we report a significant progress in the development of ZnMoO$_4$ crystal scintillators 
using deeply purified compounds (containing molybdenum with natural isotopic composition 
and enriched in $^{100}$Mo). We also present results of both aboveground and underground 
low temperature tests of new scintillating bolometers based on natural ZnMoO$_4$ and 
enriched Zn$^{100}$MoO$_4$ crystal scintillators in light of their possible application to 
next-generation 0$\nu$2$\beta$ decay experiments.

%--------------------------------------------------------------------------------------------
\section{Development of zinc molybdate based scintillating bolometers}

A precursor of the LUMINEU programme, a slightly yellow colored 313 g ZnMoO$_4$ 
sample with irregular shape, was produced from the first large volume ZnMoO$_4$ crystal 
boule grown by the low-thermal-gradient Czochralski (LTG Cz) technique 
\cite{Pav93,Gal14} in the Nikolaev Institute of Inorganic Chemistry (NIIC, Novosibirsk, 
Russia). The second sample (with mass 329 g) produced from this boule was  tested as a 
scintillating bolometer at the Gran Sasso National Laboratories (LNGS, Assergi, Italy) \cite{Bee12c}. 

Advanced ZnMoO$_4$ crystal boules with mass of $\sim$~1 kg have been produced recently at the NIIC 
by using the LTG Cz growth technique and molybdenum purified by sublimation in vacuum and double 
recrystallization from aqueous solutions \cite{Ber14}. The crystals were recrystallized to improve 
quality of the material, and two colorless ZnMoO$_4$ cylindrical samples (with size 
$\oslash$50$\times$40~mm and mass 336 and 334 g) were produced from them. Moreover, 
a zinc molybdate crystal boule (with mass 171 g) enriched in $^{100}$Mo 
to 99.5\%  was developed for the first time at the NIIC \cite{Bar14,Gri14}, and two scintillation 
elements (with mass 59 and 63 g) were cut from the boule. The enriched molybdenum was purified 
by sublimation and recrystallization from aqueous solutions. It is worth noting the high yield 
of the Zn$^{100}$MoO$_4$ crystal boule from the initial charge (84\%) and low level of total 
irrecoverable losses of enriched material (4\%) achieved in the frame of this R\&D \cite{Bar14}. 
Some coloration of the crystal (in contradiction with the practically colorless samples produced 
from natural molybdenum) can be explained by remaining traces of iron in the enriched molybdenum 
and by crystallization procedure performed only one time \cite{Bar14}.

\begin{figure}[!ht]
\centering
\subfigure[]{
  \includegraphics[width=0.4\textwidth]{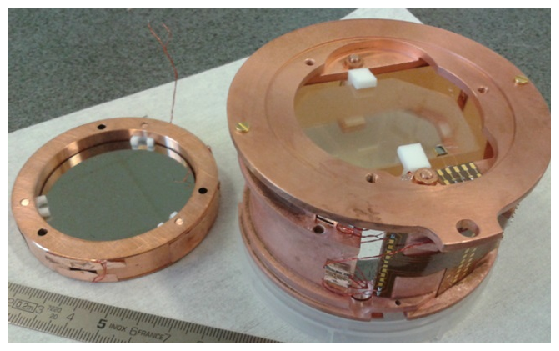}
}
\subfigure[]{
  \includegraphics[width=0.4\textwidth]{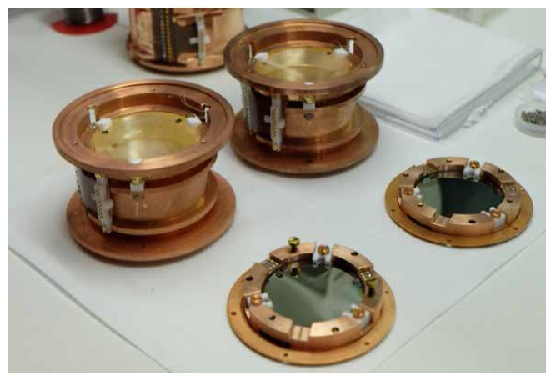}
}
\subfigure[]{
  \includegraphics[width=0.4\textwidth]{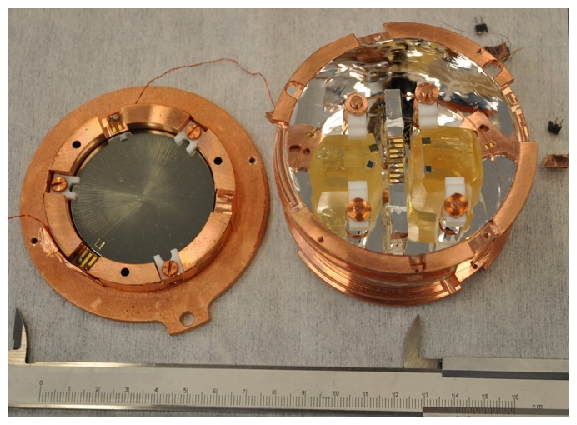}
}
\caption{Photographs of scintillating bolometers based on ultra-pure Ge photodetectors and 
ZnMoO$_4$ precursor with mass of 313 g (a),
advanced quality (see text) ZnMoO$_4$ crystals with masses of 336 and 334 g (b), and 
enriched Zn$^{100}$MoO$_4$ crystals with masses of 59 and 63~g (c).}
\label{fig:bolo}
\end{figure}

In order to construct scintillating bolometers, all the above described samples were 
held inside Copper holders by using PTFE clamps. Both Zn$^{100}$MoO$_4$ crystals 
were mounted in one Copper holder. The crystal scintillators were surrounded by a 
reflector foil (3M VM2000/2002) to improve light collection. Thin ultra-pure Ge 
wafers ($\oslash$50$\times$0.25~mm) were used for detecting scintillation light. 
The 313 g crystal was viewed by two light detectors fixed on the opposite sides. 
The ZnMoO$_4$ / Zn$^{100}$MoO$_4$ crystals and the Ge photodetectors were instrumented 
with Neutron Transmutation Doped (NTD) Ge thermistors used as temperature sensors. 
All the crystals were also assembled with an individual heating element based on a heavily-doped 
silicon meander. Such devices provide a stable resistance value and are used to inject periodically 
a certain amount of thermal energy with the aim to control and stabilize the thermal bolometric 
response. All the detector modules are shown in Fig.~\ref{fig:bolo}.
%--------------------------------------------------------------------------------------------
\section{Aboveground low temperature tests}

The 313 g ZnMoO$_4$ precursor and both Zn$^{100}$MoO$_4$ crystals were tested 
in aboveground cryogenic facilities of the Centre de Sciences 
Nucl\'{e}aires et de Sciences de la Mati\`{e}re (CSNSM, Orsay, France) 
with ``wet'' and ``dry'' $^3$He/$^4$He dilution refrigerators, respectively. 

Both cryostats are surrounded by passive shield made of low activity lead to minimize 
signals pile-up caused by environmental gamma rays due to a slow time response 
of the bolometers (hundreds millisecond). The stream data were recorded by a 16 bit ADC with a sampling 
frequency of 30 kHz and 10 kHz for natural and enriched detectors, respectively. 
The ZnMoO$_4$ precursor was operated at 17 mK during the measurements (over 38 h), 
while the Zn$^{100}$MoO$_4$ array was tested at 13.7 mK (18 h), 15 mK (5 h), 
and 19 mK (24 h) base temperatures. Both detectors were irradiated by gamma quanta from 
a weak $^{232}$Th source, while the photodetectors were calibrated with the help of 
$^{55}$Fe sources fixed close to the Ge slabs. 

The data treatment (here and below) was performed by using the optimum filtering \cite{Gat86}. 
The spectrometric performances of the precursor-based bolometer were deteriorated 
by the pile-ups effect due to considerably high counting rate $\approx$ 2.5 Hz 
(e.g. see in Table~\ref{tab:1} the energy resolution of the 2615 keV $\gamma$ peak). 
In spite of this, the test shows normal operability of the detector and allows us to 
estimate the scintillation light yield for the registered $\gamma(\beta)$ events and muons, 
as well as the possibility of particle discrimination between $\gamma(\beta)$ and 
$\alpha$ events due to the quenching of scintillation for $\alpha$ particles. 
All these data are reported in Table~\ref{tab:1}.

\begin{table*}
\caption{List of achieved performances with ZnMoO$_4$ and Zn$^{100}$MoO$_4$ detectors tested 
in aboveground and underground measurements. We report the energy resolution 
for the heat channels (FWHM --- Full Width at the Half of Maximum) estimated as filtered baseline 
and measured for $\gamma$ quanta and $\alpha$ particles of internal $^{210}$Po. We report also  
the light yield for $\gamma(\beta)$ events (LY$_{\gamma(\beta)}$) and quenching factor 
for $\alpha$ particles (QF$_{\alpha}$).}
\begin{center}
\scriptsize
\begin{tabular}{|ll|lllll|l|l|}
 \hline
\multicolumn{2}{|c|}{Detector} & \multicolumn{5}{c|}{FWHM (keV)}                    & LY$_{\gamma(\beta)}$ & QF$_{\alpha}$ \\
\cline{1-7}
 Crystal          & Mass (g) & Baseline   & $^{133}$Ba & $^{214}$Bi & $^{208}$Tl & $^{210}$Po & (keV/MeV) & ~ \\
 ~                & ~        & ~          & 356 keV    & 609 keV    & 2615 keV   & 5407 keV   & ~         & ~  \\
\hline
 ZnMoO$_4$        & 313      & 1.4(1)     & 6.4(1)     & 6(1)$^*$   & 24(2)$^*$ / 9(2)  & 19(1) & 0.77(11)$^*$ & 0.15(2)$^*$ / 0.14(1) \\
 ~                & 336      & 1.5(2)     & 6(1)       & --         & --         & 29(4)      & --   & --        \\
 ~                & 334      & 1.06(3)    & 3.8(4)     & --         & --         & 15(1)      & --   & 0.19(2)   \\
 \hline
Zn$^{100}$MoO$_4$ & 59       & 1.4(1)$^*$ & --         & 5.0(5)$^*$ & 11(3)$^*$  & --         & 1.01(11)$^*$   & $\approx 0.15^*$ \\ 
 ~                & 63       & 1.8(1)$^*$ & --         & 10(1)$^*$  & --         & --         & 0.93(11)$^*$   & $\approx 0.15^*$ \\  
\hline
\multicolumn{9}{l}{* --- results based on the aboveground measurements} \\
\end{tabular}
\end{center}
 \label{tab:1}
 \end{table*}
\normalsize

\begin{figure}[!ht]
\centering
\includegraphics[width=0.43\textwidth]{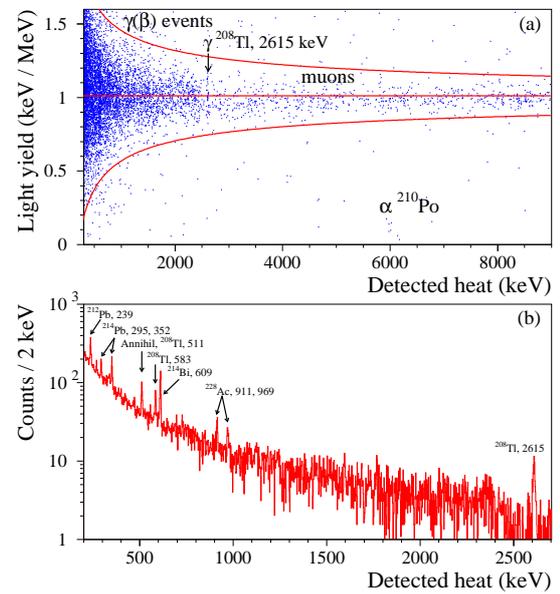}
\caption{
(a) The scatter plot of the light-to-heat signal amplitude ratio as a function 
of the heat signal amplitude accumulated at 13.7 mK 
in aboveground test with the 59 g Zn$^{100}$MoO$_4$ scintillating 
bolometer during 18 h of calibration measurements with the $^{232}$Th source. 
The visible band is related to ${\gamma(\beta)}$ events (below 2.6 MeV) and cosmic muons. 
Three sigma intervals of the light yield for the ${\gamma(\beta)}$ band are shown 
by solid red curves together with the median value. 
(b) The energy spectrum built from the data presented in the upper plot. 
The peaks observed in the energy spectrum belong to the $^{232}$Th source and environmental 
gamma's (daughters of $^{226}$Ra). The energy of marked peaks are given in keV.}
\label{fig:enr1-zmo-data}
\end{figure}

Both enriched crystals demonstrate similar performance at all the 
temperatures \cite{Bar14}. The 2-dimensional histogram obtained 
from the heat-light double read-out of the 59 g Zn$^{100}$MoO$_4$ bolometer 
at 13.7 mK is shown in Fig.~\ref{fig:enr1-zmo-data} (a). The light and the heat 
signals detected simultaneously allow to get a clear discrimination 
between $\alpha$ and ${\gamma(\beta)}$ particles.
The absence in Fig.~\ref{fig:enr1-zmo-data} (a) of peculiarities related 
with the detection of $\alpha$ events (except a small structure possibly caused by 
$^{210}$Po, as often occurs in scintillators) indicates on 
encouraging radiopurity of the tested enriched crystals. Good spectrometric 
properties of the enriched detectors, even at aboveground conditions, 
are well visible from Fig.~\ref{fig:enr1-zmo-data} (b), while some further 
information about their performances is presented in Table~\ref{tab:1}.

%--------------------------------------------------------------------------------------------
\section{Underground cryogenic measurements}

The 313 g detector was moved deep underground ($\approx$4800 m w.e.) to the 
Modane Underground Laboratory (Laboratoire Souterrain de Modane, LSM, France) 
and tested during the EDELWEISS-III commissioning runs. The ZnMoO$_4$ bolometer 
together with fifteen ultra-pure Ge detectors (0.8 kg each) fully covered with 
interleaved electrodes (FID) were installed inside the $^3$He/$^4$He inverted dilution 
refrigerator with a large experimental volume (50 $l$) \cite{Arm11}. 
The EDELWEISS set-up, located inside a clean room (ISO Class 4) and supplied by 
deradonized ($\approx$30 mBq/m$^3$) air flow, is surrounded by a massive shield made of 
low background lead (20 cm thick) and polyethylene (50 cm). The set-up is surrounded 
by a 5 cm thick plastic scintillator muon veto (95\% coverage), and equipped by 
neutron and radon counters. 

The triggered signals were recorded by a 14 bit ADC in 2 s window with 2 kHz sampling rate 
(the half of the window contains the baseline data). The base temperature was stabilized 
around 19 mK. One light detector was very sensitive to microphonic noise and could not be 
used for measurements. The energy scale of the ZnMoO$_4$ detector has been measured in 
calibration runs with $^{133}$Ba and $^{232}$Th $\gamma$ sources, performed  
over 546 h and 70 h, respectively. The background data were accumulated over 305~h.

\begin{figure}[!ht]
\centering
\includegraphics[width=0.43\textwidth]{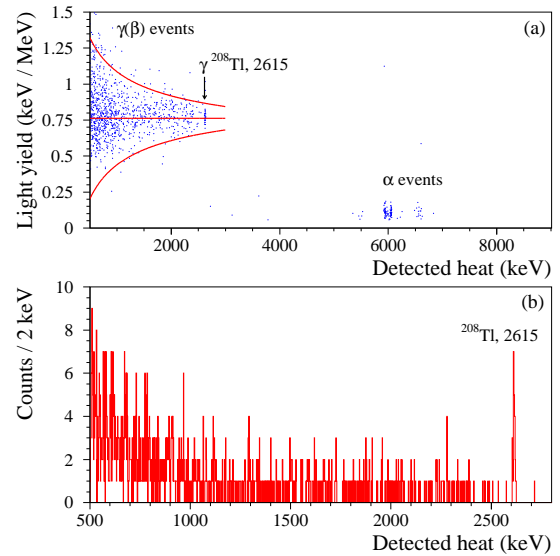}
\caption{
(a) Plot reporting the light-to-heat signal amplitude ratio as a function 
of the heat signal amplitude for the 313 g ZnMoO$_4$ detector installed 
in the EDELWEISS set-up. The detector was cooled down to 19 mK and irradiated over 51 h 
by $\gamma$ quanta from the $^{232}$Th source. Two visible bands correspond 
to ${\gamma(\beta)}$ events and $\alpha$ particles. 
The positions of the $\alpha$ events are shifted from the nominal values due to thermal 
energy overestimation for $\alpha$ particles in case of using calibration data for 
$\gamma$'s. Three sigma intervals of the light yield for the ${\gamma(\beta)}$ 
band and its median value are drawn. 
(b) The energy spectrum of the $^{232}$Th source measured by the 313 g ZnMoO$_4$ 
scintillating bolometer during 51 h of underground cryogenic run.}
\label{fig:nat0-zmo-data}
\end{figure}

The powerful discrimination capability achieved with the 313 g ZnMoO$_4$ 
scintillating bolometer is well illustrated in Fig.~\ref{fig:nat0-zmo-data} (a),  
which shows a full separation of ${\gamma(\beta)}$-induced events from populations 
of $\alpha$ particles caused by trace impurity by radionuclides from U/Th chains 
(mainly, $^{210}$Po, see below). The energy spectrum accumulated with the $^{232}$Th 
gamma source (see Fig.~\ref{fig:nat0-zmo-data} (b)) demonstrates high 
spectrometric properties of the detector. An overview of the detector's 
performances during underground measurements is given in Table~\ref{tab:1}. 

After completing the EDELWEISS-III commissioning runs, other two ZnMoO$_4$-based 
scintillating bolometers ($\oslash 50 \times 40$ mm) and the Zn$^{100}$MoO$_4$ array 
together with 36 FID Ge detectors were assembled. The EDELWEISS set-up was also upgraded: 
a) a  polyethylene shield at the 1 K plate was added; 
b) new ultra radiopure NOSV Copper \cite{Lau04} screens were installed; 
c) all detectors were provided with individual low background Copper-Kapton cables. 
In addition, a pulser system to assist to the calibration of the thermal response 
of the ZnMoO$_4$ / Zn$^{100}$MoO$_4$ detectors will be implemented soon.

\begin{figure}[!ht]
\centering
\includegraphics[width=0.43\textwidth]{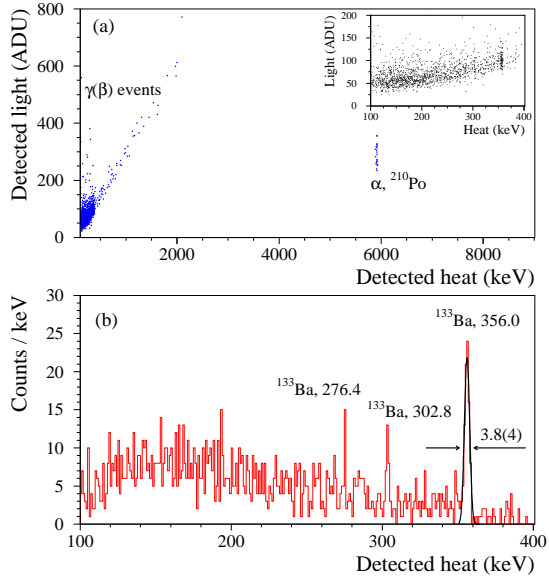}
\caption{
(a) Scatter plot of the light versus the heat signals measured by 
the 334 g ZnMoO$_4$ scintillating bolometer in a 15 h calibration run with the 
$^{133}$Ba gamma source in the EDELWEISS set-up. A cluster of events located 
far from the ${\gamma(\beta)}$ population corresponds to $\alpha$ particles 
of $^{210}$Po. The data for the light channel are presented in ADU 
(Analogue-to-Digital Unit). 
(Insert) Part of the scatter plot corresponding to the energy range 
of the used source.
(b) The energy spectrum of the $^{133}$Ba source measured over 15 h by the 
334 g ZnMoO$_4$ scintillating bolometer.}
\label{fig:nat2-zmo-data}
\end{figure}

After the upgrade of the set-up the data are recorded by a 16 bit ADC with 1 kHz sampling rate 
(the length of pulse profile is 2 s with the half of the window for the baseline data).
The working temperature is stabilized at 18 mK. The energy scale of the detectors was 
measured with the $^{133}$Ba gamma source (the measurements with the $^{232}$Th source 
are foreseen). 

The set-up is still under optimization, especially as far as 
the control of the vibration-induced noise is concerned. Therefore, we discuss here, 
as an illustrative example, only the results achieved with the 334 g natural ZnMoO$_4$ 
scintillating bolometer. This detector exhibits full $\alpha$/${\gamma(\beta)}$ separation, 
as shown in Fig.~\ref{fig:nat2-zmo-data} (a), as well as excellent spectrometric properties, 
as demonstrated in Fig.~\ref{fig:nat2-zmo-data} (b). Other relevant information 
about performances of $\oslash 50 \times 40$ mm  ZnMoO$_4$ detectors 
are reported in Table~\ref{tab:1}. 

%--------------------------------------------------------------------------------------------
\section{Radiopurity of ZnMoO$_4$ and Zn$^{100}$MoO$_4$ crystals}

\begin{figure}[!ht]
\centering
\includegraphics[width=0.43\textwidth]{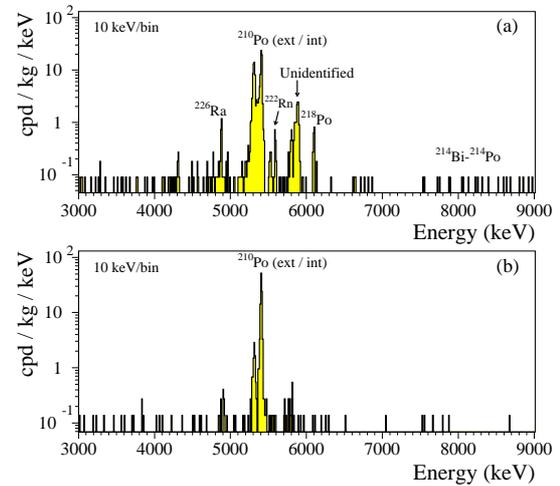}
\caption{The $\alpha$ spectra collected in the low background measurements 
in the EDELWEISS set-up with the ZnMoO$_4$ scintillating bolometers based on the 313 g 
precursor (a) and the 334 g advanced sample (b) operated over 851 h and 527 h, 
respectively. The origin of the $\alpha$ events providing the highest rate are indicated.}
\label{fig:nat-zmo-alpha}
\end{figure}

The radiopurity level of the ZnMoO$_4$ crystals was estimated by analysis of 
the $\alpha$ events selected from the underground runs, while the data of  
the aboveground measurements were used in case of the Zn$^{100}$MoO$_4$ 
samples. The position of the 5.4 MeV $\alpha$ peak of the internal $^{210}$Po, 
clearly visible in the data for the natural crystals, was used to stabilize 
the thermal response of the detectors. For instance, the spectra of the $\alpha$ 
events registered by the detectors based on 313 g (a) and 334 g (b) ZnMoO$_4$ 
crystals over 851 h and 527 h, respectively, are shown in 
Fig.~\ref{fig:nat-zmo-alpha}.  

The crystals are slightly polluted by $^{210}$Po detected through 5.4 MeV $\alpha$ 
peak confirming a broken equilibrium in the radioactive chain. $^{226}$Ra 
(and its daughters $^{222}$Rn, $^{218}$Po, and $^{214}$Bi-$^{214}$Po events), 
and $^{228}$Th (with daughter $^{224}$Ra\footnote{Taking into account a short 
half-life of $^{216}$Po ($\approx$145 ms), which is comparable with the time  
response of the 313 g detector (hundreds ms), subsequent $\alpha$ decays of 
$^{220}$Rn-$^{216}$Po give pile-ups and therefore were discarded from the data 
by the pulse-shape analysis.}) were detected in the 313 g crystal, 
while the ZnMoO$_4$ scintillators produced by recrystallization have shown 
a much better level of radiopurity, particularly in $^{226}$Ra. 
It is also evident a higher surface contamination by $^{210}$Po of the 313 g 
crystal or/and of the bolometer components close to it (a peak at 5.3 MeV 
corresponds to $E_{\alpha}$ of $^{210}$Po). In addition, excess counts around 
5.8 MeV also indicate a possible surface contamination but its origin has not 
been identified. 

The activity of internal $^{210}$Po was derived from the fit of the 5.4 MeV peak,
while 3$\sigma$ intervals (according to the energy resolution of the internal 
$^{210}$Po --- see Table~\ref{tab:1}) centered at the $Q_{\alpha}$ value 
were used for the calculation of the area of the peaks of other radionuclides 
from U/Th chains. The background contribution was evaluated in 
two energy regions (3.3--4 and 4.35--4.7 MeV) with a flat $\alpha$ continuum 
in which no peaks are expected. The number of counts excluded with 90\% C.L. 
were calculated by using the Feldman-Cousins procedure \cite{Fel98}. 

%\footnotesize
\begin{table}[!ht]
%\begin{table*}[!ht]
\caption{Radioactive contamination of the ZnMoO$_4$ and Zn$^{100}$MoO$_4$ crystals 
tested as scintillating bolometers in aboveground and underground conditions. The mass 
of the crystals and the total time of the accumulated data are also presented.  
The results for the large mass ZnMoO$_4$ crystal which was operated as the 
scintillating bolometer at the LNGS (Italy) \cite{Bee12c} are given for comparison. 
The uncertainties are given with 68\% C.L., while all the limits are at 90\% C.L.}
\scriptsize
\begin{center}
\begin{tabular}{|l|ll|lll|l|}
 \hline
 Nuclide    & \multicolumn{6}{c|}{Activity (mBq/kg)} \\
\cline{2-7}
 ~         & \multicolumn{2}{c|}{Zn$^{100}$MoO$_4$} & \multicolumn{4}{c|}{ZnMoO$_4$}  \\
\cline{2-7}
 ~          & 59  g      & 63 g       & 336 g       & 334 g       & 313 g       & 329 g \cite{Bee12c} \\
\cline{2-7}
 ~          & 42 h       & 42 h       & 291 h       & 527 h       & 851 h       & 524 h \\
 \hline
$^{228}$Th  & $\leq0.25$ & $\leq0.21$ & $\leq0.024$ & $\leq0.007$ & 0.010(3)    & $\leq0.006$ \\
\hline
$^{238}$U   & $\leq0.26$ & $\leq0.21$ & $\leq0.008$ & $\leq0.002$ & $\leq0.008$ & $\leq0.006$ \\
$^{226}$Ra  & $\leq0.26$ & $\leq0.31$ & $\leq0.021$ & $\leq0.009$ & 0.26(5)     & 0.27(6)   \\
$^{210}$Po  & 0.9(3)     & 1.1(3)     & 0.94(5)     & 1.02(7)     & 0.62(3)     & 0.70(3) \\
 \hline
\end{tabular}
\end{center}
 \label{tab:2}
 %\end{table*}
 \end{table}
\normalsize

Data (or limits) on radioactive contamination of the ZnMoO$_4$ and Zn$^{100}$MoO$_4$ 
scintillators are summarized in Table \ref{tab:2}, where the results for another 
ZnMoO$_4$ sample, produced from the same boule as the 313 g crystal was, 
are presented for comparison. As it is seen from Table \ref{tab:2}, 
the improved purification and crystallization procedure adopted 
for the LUMINEU crystals of 334 and 336 g has lead to a significant reduction of 
the internal contamination, especially for $^{226}$Ra which is not detectable 
now while it was clearly present in both precursor crystals (313 and 329 g). In particular, 
the radiopurity levels ( $\leq$0.01 mBq/kg) achieved for $^{228}$Th and 
$^{226}$Ra are fully compatible with next-generation 0$\nu$2$\beta$ experiments capable 
to explore the inverted hierarchy region of the neutrino mass pattern \cite{Bee12,Art14}. 

%--------------------------------------------------------------------------------------------
\section{Conclusions}

A significant progress is achieved in development of ZnMoO$_4$ crystal scintillators 
for the LUMINEU project. Large volume crystal boules ($\sim$ 1 kg each) were grown by the 
low-thermal-gradient Czochralski technique from deeply purified molybdenum. 
A Zn$^{100}$MoO$_4$ crystal boule with a mass of 0.17 kg was produced from enriched 
$^{100}$Mo (to 99.5\%) for the first time. Three natural ($\sim$ 0.3 kg) and 
two enriched ($\sim$ 0.06 kg) scintillation elements were produced for low temperature 
studies. Production of large volume Zn$^{100}$MoO$_4$ 
crystal scintillators from enriched $^{100}$Mo is in progress. 

The cryogenic scintillating bolometric tests of the natural and enriched crystals 
showed a high performance of the detectors. The deep purification of molybdenum and 
recrystallization significantly improve the radioactive contamination of 
ZnMoO$_4$ crystals by $^{228}$Th and $^{226}$Ra to the level of $\leq$0.01 mBq/kg 
requested by the LUMINEU project. 

The results of this study clarify the excellent prospects of ZnMoO$_4$ 
scintillating bolometers for the next generation $0\nu 2\beta$ experiments aiming to 
approach the inverted hierarchy region of the neutrino mass pattern.

%--------------------------------------------------------------------------------------------
\section{Acknowledgments}

The development of ZnMoO$_4$ scintillating bolometers is part
of the LUMINEU programme, a project receiving funds from the Agence
Nationale de la Recherche (ANR, France). The work was supported in part
by the project ``Cryogenic detector to search for neutrinoless
double beta decay of molybdenum'' in the framework of the Programme
``Dnipro'' based on Ukraine-France Agreement on Cultural, Scientific
and Technological Cooperation. The aboveground bolometric tests were 
assisted by ISOTTA, a project receiving funds from the ASPERA 2nd Common
Call dedicated to R\&D activities. D.V.~Poda was supported by the 
P2IO LabEx (ANR-10-LABX-0038) in the framework ``Investissements d'Avenir'' 
(ANR-11-IDEX-0003-01) managed by the ANR (France).


\end{document}